# Monolithic 3D FPGAs Utilizing Back-End-of-Line Configuration Memories

Faaiq Waqar[1*], Jiahao Zhang[2], Anni Lu[1], Zifan He[2], Jason Cong[2], Shimeng Yu[1*]
[1]Georgia Institute of Technology, Atlanta, GA; [2]University of California – Los Angeles, Los Angeles, CA
[*]Email: faaiq.waqar@gatech.edu, shimeng.yu@ece.gatech.edu

*Abstract*— This work presents a novel monolithic 3D (M3D) FPGA architecture that leverages stackable back-end-of-line (BEOL) transistors to implement configuration memory and pass gates, significantly improving area, latency, and power efficiency. By integrating n-type (W-doped $In_2O_3$) and p-type (SnO) amorphous oxide semiconductor (AOS) transistors in the BEOL, Si SRAM configuration bits are substituted with a less leaky equivalent that can be programmed at logic-compatible voltages. BEOL-compatible AOS transistors are currently under extensive research and development in the device community, with investment by leading foundries, from which reported data is used to develop robust physics-based models in TCAD that enable circuit design. The use of AOS pass gates reduces the overhead of reconfigurable circuits by mapping FPGA switch block (SB) and connection block (CB) matrices above configurable logic blocks (CLBs), thereby increasing the proximity of logic elements and reducing latency. By interfacing with the latest Verilog-to-Routing (VTR) suite, an AOS-based M3D FPGA implemented in 7 nm technology is demonstrated with 3.4× lower area–time squared product ($AT^2$), 27% lower critical path latency, and 26% lower reconfigurable routing block power on benchmarks including hyperdimensional computing and large language models (LLMs).

## I. Introduction

Over the past three decades, field-programmable gate arrays (FPGAs) have become increasingly popular in domains such as telecommunications (packet processing and network function virtualization) [1], system-on-chip development (prototyping and verification) [2], and hardware acceleration for AI/ML workloads at the edge and in the data center [3]. The advantage of FPGAs lies in their off-the-shelf reconfigurability, enabling the implementation of custom circuit designs using hardware description languages (HDLs) or high-level synthesis (HLS), thereby circumventing the substantial non-recurring engineering (NRE) costs—such as physical design, layout, and fabrication—associated with application-specific integrated circuit (ASIC) development. This adaptability extends their operational longevity in rapidly evolving markets where fixed-function ASICs quickly become obsolete. FPGAs can execute custom applications at over 10× lower power and with >3× runtime reduction compared to CPU implementations [4]. However, although modern FPGAs incorporate hardened macros such as RAMs, processor subsystems, and digital signal processing (DSP) units, designs implemented on an FPGA are still 9× larger and 3-6× slower than an equivalent built on an ASIC [5].

The principal cause of the power, performance, and area (PPA) disparity between FPGAs and ASICs is a byproduct of their key advantage: reconfigurability. Configuration memories implemented using SRAM enable/disable logical and signal propagation functions to emulate the data path in an ASIC. Modern FPGAs require a substantial number (~2,000-5,000) of configuration bits per tile (Section 2A). The low density and high static power consumption of SRAM cells significantly limit the density of FPGA designs; configuration memories can occupy >50% of a tile's area and account for ~12% of the total static power [6]. Additionally, the routing fabric comprises extensive networks of crossbars, multiplexers, buffers, and wires that dominate the dynamic (~75%) and static (~78%) power consumption [7]. We observe that reductions to the routing and reconfiguration overhead have significant implications on the PPA of FPGAs.

Monolithic 3D (M3D) integrated circuits, enabled by innovations in low-temperature materials processing, permit the use of multiple active tiers on a single substrate by building transistors in the back-end-of-line (BEOL). Among the most promising emerging transistor candidates are amorphous oxide semiconductor (AOS) transistors, owing in part to their commercial adoption in transparent thin-film channels for active display technology [8]. Beyond their BEOL compatibility and stackability, AOS transistors have ultra-low leakage ($<10^{-15}$-$10^{-18}$ μA/μm) [9], high $I_{on}/I_{off}$ ratio ($>4×10^9$), and moderate electron mobility (~15-20 $cm^2/V·s$) [10]. AOS transistors have found themselves at the forefront of research on charge-based memories [11], hybrid M3D standard cells [12], and BEOL power delivery [13]. Although vast literature exists on the applications of AOS transistors, their candidacy as a fully BEOL-compatible SRAM substitute has been largely unexplored due to the lack of a strong p-type AOS substitute with comparable hole-mobility which by translation degrades the dynamic access speed critical for SRAM-based register and cache memory. However, in FPGAs, SRAM configuration bits are stationary at runtime. Thus, the preferable criterion in a configuration SRAM bit cell is governed by its device stability, static power, and footprint, on which front BEOL-compatible AOS SRAMs can improve upon the Si counterpart [14].

This paper presents a design-space analysis of a novel M3D FPGA architecture with AOS transistor-based configuration SRAMs and multiplexed routing structures. Our proposed architecture bypasses the need for high-voltage conversion and delivery for programming (a density/reliability bottleneck in prior work on FPGAs employing emerging devices). To enable the precise quantitative study of an M3D FPGA with AOS device integration and its advantages over CMOS FPGAs in 7 nm technology, we develop an evaluation flow based on robust compact transistor models, a custom M3D-compatible version of COFFE [15] and Verilog-to-Routing (VTR) [16] to appraise PPA improvements on benchmarks targeting hyperdimensional computing and natural language processing (NLP).

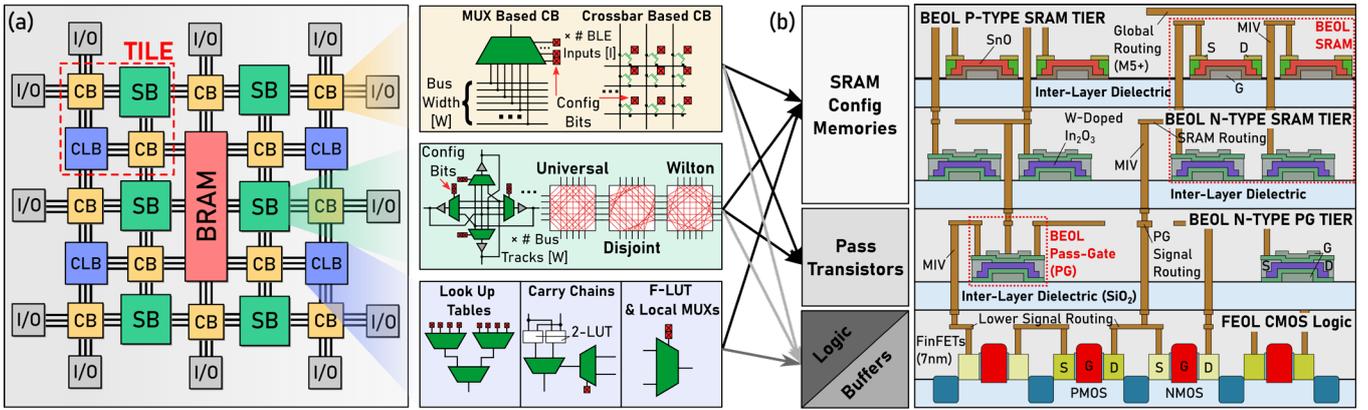

Fig. 1. (a) Island-based FPGA organization with breakdown of reconfigurable logic (CLB) and routing (Switch (SB), Connection (CB)) blocks. Extensive use of configuration bits and pass-transistors shown in red and green (b) proposed M3D FPGA organization of FEOL/BEOL devices and routing structures.

## II. BACKGROUND AND RELATED WORK

### A. Conventional FPGA

Fig. 1a illustrates a conventional island-based FPGA design, composed of an array of tiles—modular units containing a Configurable Logic Block (CLB) and routing resources in the form of a Switch Block (SB) and pairs of Connection Blocks (CBs). Each CLB houses a cluster of $N$ Basic Logic Elements (BLEs), as shown in Fig. 2a. The BLE datapath comprises a look-up table (LUT) that maps a $K$-input logical function, where inputs are fed to the selection pins of the LUT multiplexer. LUTs are interfaced with a CLB input mux grid to select operation inputs, flip-flops (FFs) to hold the logical expression's output, and, at times, Fracturable LUT (F-LUT) multiplexers to break down LUTs for higher utilization [17]. At the output, feedback multiplexers feed data recursively back to the input while carry chains sometimes expedite arithmetic within BLEs [18]. A global bus consisting of horizontal and vertical routing channels with multiple tracks ($W$) that span several tiles ($L$) connects to CLB input pins ($I$) through CBs, using either crossbar or mux-based interconnects. SBs are positioned at the intersections of routing channels to direct signals across various paths using routing switches, the number and placement of which depend on the topology (Fig. 1a) and the block's connection flexibility ($F_s$). All three principal blocks within a tile extensively utilize configuration bits fed into selection pins (and, thus, transistor gates) for routing and I/O selection and as inputs for LUT multiplexers. These interconnected blocks form a datapath, depicted in Fig. 2b, the longest of which determines the critical path delay (CPD) within a layout after place and route (P&R).

### B. Amorphous Oxide Semiconductor Transistors

Amorphous oxide semiconductors (AOS) such as the indium oxide ($In_2O_3$) family are a class of materials that exhibit moderate electron mobility (~10–20 cm²/V·s) and can be fabricated at BEOL compatible temperatures (<400 ºC). In n-type AOS materials, conduction is primarily governed by donor defects such as oxygen vacancies. To suppress the formation of oxygen vacancies and enhance the stability of the film, dopants with high oxygen affinity, such as gallium (Ga), zinc (Zn), tin (Sn), or tungsten (W), are incorporated. Leading foundries and research institutes such as TSMC, Samsung, and IMEC are actively researching the enhancement of AOS transistors [14],[19]-[20]. Among various n-type AOS devices, IWO was selected for this study because of its balanced characteristics of high on-state current density and threshold voltage stability [21].

Achieving p-type conduction in AOS is more challenging due to the localized nature of the valence band, involving acceptor defects like cation vacancies that facilitate hole conduction. Recently, TSMC reported a p-type tin-oxide (SnO) transistor with hole mobility as high as ~2 cm²/V·s and a record $I_{on}$ of 10~20 μA/μm [14]. Proposed AOS devices for M3D integration candidacy have been demonstrated down to channel lengths ($L_{ch}$) of 10 nm in n-type IGZO [20], 20 nm n-type IWO [22], and 25 nm in p-type SnO [14]. The low-temperature fabrication process permits AOS transistors to be built above the Si logic layer without damaging the underlying devices, allowing active logic tiers to be stacked among the BEOL interconnect levels (Fig. 1b).

### C. FPGAs using Emerging Memories and Devices

Prior works targeting emerging device integration for FPGA optimization fall into one of three categories: (1) the substitution of configuration memories and/or pass gates, (2) the truncation of multiplexing circuits, and (3) the expansion of block RAM (BRAM). This study belongs to (1), in which extensive work has been conducted on the use of resistive switching mechanisms in RRAM [23]-[25], FeFET [26], NEM relay [7],[27]-[28], and PCM [29]. Among RRAM-based FPGA studies, [23] proposes the replacement of pass-gates with RRAM in SBs/CBs and the use of 2-RRAM divider configuration memories to improve footprint 2-3× and reduce power by 20%, FPGA-RPI [24] develops routing/programming methodology and a BEOL place and route EDA flow with an adaptive buffering scheme for RRAM SBs/CBs that can reduce the programmable mesh area by 96% and improves performance by 55%, and [25] tapes-out a 1T-2R divider-based BEOL configuration memory over 289 tiles in 180 nm node reducing area by 57%. In [26], FeFET-based routing elements and current-based LUTs are proposed, reducing FPGA footprint by 8% and improving power-delay-product (PDP) by 5.8-16×, although the current-based LUT readout non-trivially increases static power consumption. Notable studies on NEM relay-based FPGAs include [27], which proposes relay routing structures and divider-based configuration memories to cut the area by 43.6% and latency by 28%, [28] that extends this further with face-to-face heterogeneous 3D integration to reduce latency by 42% over 2D CMOS, and [7] which demonstrates the functional operation of a 2-by-2 NEM relay for BEOL CBs. However, a challenge of employing the proposed emerging non-volatile memories (eNVMs) is the reliance on high-voltage programming, which necessitates the use of external circuitry and programming grid with high-voltage tolerance (such as thick-oxide I/O transistors) degrading density and reliability. Additionally, surface forces dominate over inertial forces in scaled NEM relays, hindering programming reliability. Strategic tradeoffs between prospective devices are further investigated in Section 3B.

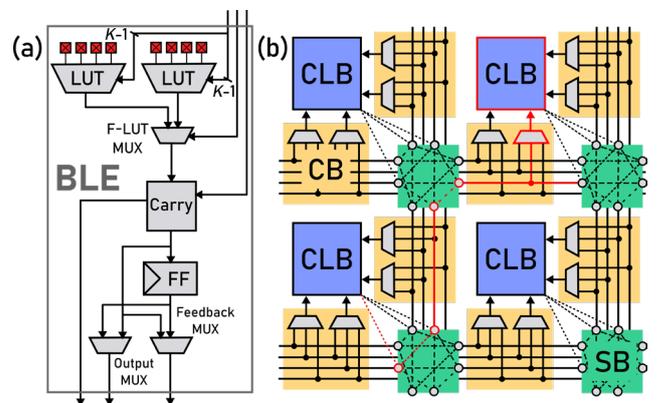

Fig. 2. (a) Composition of a basic logic element (BLE) in a CLB (b) Example of a critical path (red) in an island-based FPGA design, CLBs use LUTs to emulate logic and utilize SB and CB blocks to connect to one another. The placement of logical elements and routing paths is determined through a simulated annealing algorithm during place and route (P&R)

### III. BEOL COMPATIBLE CONFIGURATION BITS

#### A. Memory Cell Topology and Design

As noted in Section 1, configuration SRAM cells, distributed widely throughout CLBs, SBs, and CBs, can be quantitatively assessed by their ability to drive sufficient voltage and by their static power. For example, Xilinx transitioned to using larger middle-thickness oxide transistors for SRAM to lower the static power consumption in 40-45 nm technology nodes [30]. A cross-sectional view of our M3D placement of AOS SRAM cells is depicted in the upper tiers of Fig. 1b, above the lower metallization tiers (M1-M4). To minimize the footprint of each cell, two tiers are used to construct the AOS SRAMs, the upper for PMOS and the lower for NMOS. The storage node (Q) of the SRAM cell must connect to the local interconnect below to be accessed by front-end-of-line (FEOL) logic, which requires the signal to propagate through to M1 or M2. This transmission is done using monolithic interlayer vias (MIVs).

To evaluate the proposed AOS SRAM design, short channel ($L_{ch}$=25 nm) machine learning-assisted compact models [31] are utilized for calibrating the TCAD or experimental data reported from double-gated (DG) IWO (n-type) [22] and back-gated (BG) SnO (p-type) devices [14], which is then validated in SPICE simulation. The BG SnO FET compact model is calibrated directly from experimental data of a 25 nm channel length device. To determine the behavior of channel length scaling in DG IWO FETs with dual-gate operation, we model the structural changes made to an experimentally calibrated DG IWO FET in Sentaurus TCAD at 50 nm channel length, which captures the short channel effects of the AOS transistor. Then, the ML-assisted compact model is developed from Id-Vg, Id-Vd, and C-V data extracted from TCAD. Tile design parameters used to benchmark FPGA performance with/without AOS integration are set to {$K, N, I, W, L, F_s$} = {6, 10, 40, 150, 8, 3}.

#### B. Power, Stability, and Programmability

The bit cell layout of the proposed AOS SRAM cell is depicted in Fig. 3b. To minimize the AOS SRAM footprint, a bit cell is constructed in 2-tiers, where the inverter and feedback connections are made through pairs of MIVs, and separate metallization tiers are used for the WL and rail voltage connections. Fig. 3c depicts findings for the differences in static power between AOS and 7 nm Si-based (minimum-sized ASAP7 FinFET) configuration SRAMs. A 2:1:1 PU:PD:PG structure with a minimum width of 50 nm is used for the AOS SRAM design. When used at the same overdrive voltage ($V_{SRAM}$) of 0.8 V, AOS SRAMs can reduce static power consumption by 60.1% per bit cell over their CMOS counterpart, lowering the cumulative static power of configuration memories to ~5.3% of all static power on-chip. We compare the programming speed and voltage of both SRAMs and

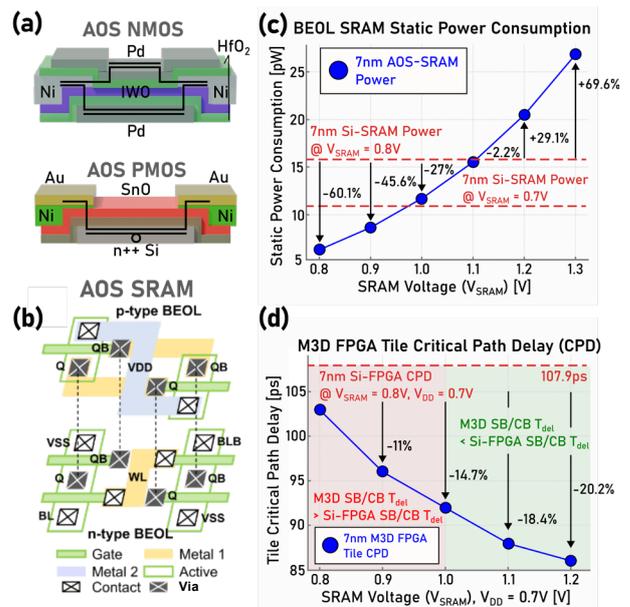

Fig. 3. (a) Device structure of double-gated (DG) n-type IWO and back-gated (BG) p-type SnO devices (b) 2-tier layout of BEOL compatible AOS SRAM (c) Comparison of AOS and CMOS SRAM static power as a function of $V_{SRAM}$ (d) Comparison of conventional and M3D FPGA tile CPD as a function of $V_{SRAM}$

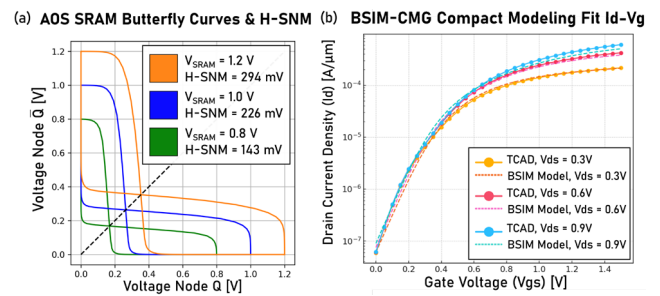

Fig. 4. (a) Butterfly curve and corresponding H-SNM of AOS SRAM vs. $V_{SRAM}$ (b) BSIM-CMG model fitting for AOS NMOS-based pass-gate

TABLE I
Pass-Device Comparison Proposed for FPGA Reconfiguration

| | RRAM [23]-[25] | NEM [27]-[28] | Si FeFET [26] | AOS FET 7 nm | Si FET 7 nm |
|---|---|---|---|---|---|
| $R_{on}/R_{off}$ | $10^3$-$10^6$ | $>10^9$ | $10^6$-$10^7$ | $10^9$ (PG) | $10^7$ (PG) |
| $V_{PRGM}$ | 3.3 V | 0.8 V* | 5 V | 0.7 V (SRAM) | 0.7 V (SRAM) |
| $T_{PRGM}$ | 5 ns | <10 ns | 2 ns | 440 ps | 17 ps |
| Area | Small (BEOL) | Large (BEOL) | Small (FEOL) | Large (BEOL) | Large (FEOL) |
| Pass-Disturb | Medium | High | Medium | Low | Low |

* Projection when scaled to 22 nm technology node metallization pitch

configuration memories proposed for SRAM substitution in prior works (Table I). Although the programming time of AOS SRAM bits is slower than that of conventional Si SRAM by a factor of 26×, the cell can still be sufficiently written using WL and BL pulses that do not need to exceed $V_{DD}$ at 7 nm (0.7 V) thanks to the restorative feedback in SRAMs cross-coupled inverter structure, even when $V_{SRAM} > V_{DD}$. This eliminates the need for a programming grid for high-voltage applications, improving density and reliability while maintaining a programming speed 4-10× shorter than other eNVM technologies proposed for FPGA reconfigurable structures. AOS SRAMs demonstrate a sense noise margin (H-SNM) of 226 mV at 1.0 V, compared to 339 mV at 0.8 V in CMOS (Fig. 4a). The leftward shift of the metastable point and, thus, lower H-SNM arises due to a much weaker AOS PMOS with negative $V_{TH}$; however, an H-SNM >200 mV is sufficient tolerate noise injected at the storage node, such as that due to capacitive coupling at the pass-gate.

### IV. BEOL COMPATIBLE PASS GATES FOR ROUTING

#### A. Equivalent Circuit Model and Topology

Fig. 5a shows that configuration memory substitution alone reduces the tile footprint by ~49%. However, there is considerable room for improvement as SBs and CBs still consume 17.6% and 13.8% of the footprint, respectively. Owing to the observation that the composure of routing blocks is almost entirely made from mux and/or crossbar structures consisting of either pass or transmission gates, the insertion of a tier of n-type IWO pass gate-based SB and CB above CLBs (Fig. 1b) could significantly reduce the cumulative footprint, which we refer to as a true M3D FPGA. The composition of this proposed M3D FPGA (with both BEOL configuration memories and pass gates) is also shown in Fig. 5a, demonstrating ~59% area reduction over the conventional CMOS tile. Furthermore, we sweep tile parameters that strongly influence tile area (Fig. 5b), demonstrating that the multi-tier M3D design improves the footprint scalability of logic complexity in CLBs. However, in tiles with large routing bus widths, congestion in the BEOL can limit footprint reduction.

#### B. Timing, Area, and Power

To measure the performance and power reductions of AOS routing pass gate (PG) structure, we modify the DG IWO FET TCAD structure to capture the device physics under 7 nm design rule ($L_{ch}$=18 nm) and model it with BSIM-CMG for SPICE simulation (Fig. 4b). Base logic input signals and buffers are connected to AOS pass gates through MIVs, for which we use parasitics of 0.18 fF and 96 Ω per MIV [32]. We assume the implementation of BEOL pass gates above the signal routing layers for CLB logic and buffers (M1-M2) and dedicate a separate signal routing tier for PGs (M3) using ASAP7 M1-M3 parasitics (131.2 Ω/μm, 0.23 fF/μm).

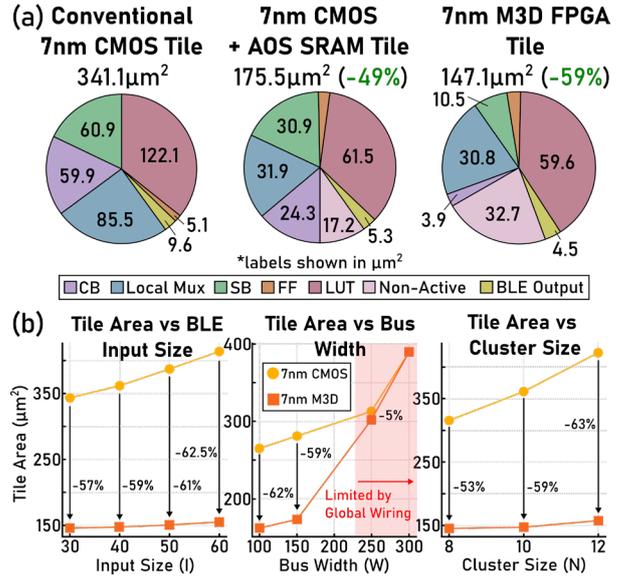

Fig. 5. (a) Composition of 7 nm CMOS, CMOS +AOS SRAM, and M3D FPGA designs (b) Parametric area comparison of CMOS and M3D tiles

In Fig. 3d, we compare the tile level critical path delay (CPD) in CMOS and M3D FPGAs as a function of the SRAM voltage ($V_{SRAM}$) under specified design parameters modeled based on the Xilinx-7 architecture. Cumulative critical path delay is reduced at the same $V_{SRAM}$ (0.8 V) by ~8% thanks to shorter routing/load RC. However, this reduction is limited by the lower drive current and higher source/drain capacitance found in AOS transistors due to the large overlap regions and lower channel electron mobility [11]. We find that overdriving the AOS pass gates to/beyond 1.0 V also improves SB/CB performance while reducing the configuration memory static power by 27%. With this in mind, we consider the split of the rail voltages of SRAM configuration bits between those set for CLBs and those used to drive AOS pass gates in SBs and CBs ($V_{SRAM}$, $V_{SRAM-SCB}$).

Buffers and the reconfigurable mesh are the highest patrons of power consumption in FPGAs, consuming ~82% of static and ~68% of dynamic system power [7]. We observe that a disadvantage of using pass-gates over transmission gates in CMOS is their inability to transmit a signal in full swing, limited by the threshold voltage of the pass device. This is supplemented by a level-restorer, which pulls up the voltage to digital-circuit readable levels. However, restorers increase power and latency. Given the larger bandgap in AOS materials, it is possible to exceed $V_{SRAM}$ voltages used on CMOS pass gates without severely degrading the reliability. A $V_{SRAM-SCB} \geq V_{DD} + V_{TH}$, such as 1.2 V, removes the need for a level-restorer while increasing SRAM static power consumption by only 29.1% in AOS SB and CB configuration bits over CMOS. However, this increase is exceeded by the cumulative power reduction in the M3D design. To evaluate this modification, we test SB and CB structures made of AOS and CMOS devices in SPICE (Fig. 6a). The average power at 250MHz is given by:

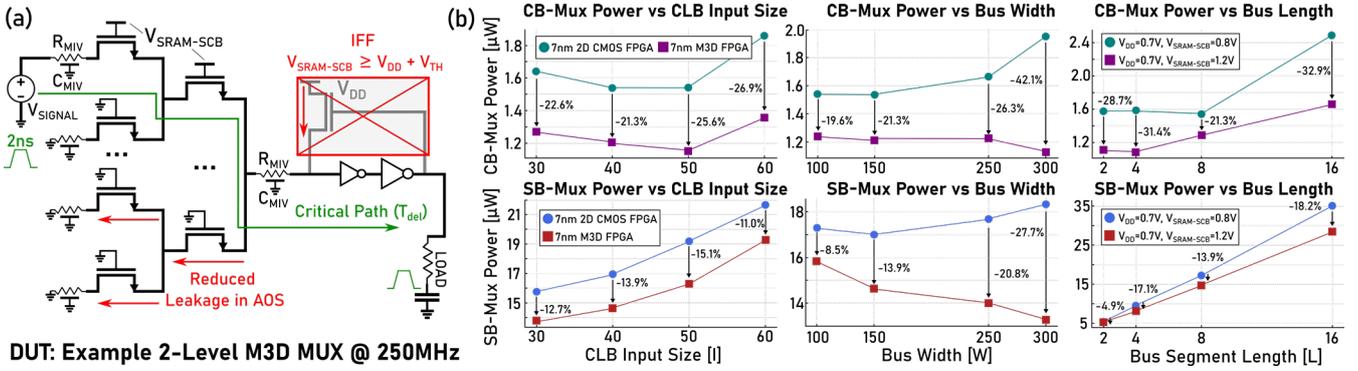

Fig. 6. (a) Design under test (DUT) for an example 4:1 mux with overdriven IWO pass gates, enabling removal of level-restorers (b) parametric sweep of bus parameters vs. SB/CB mux power consumption. Level-restorer removal and low leakage in AOS devices improve the efficiency of reconfigurable routing.

$$P_{avg} = \int_0^{4ns} V_{dd} \times I_{src} dt \Big/ 4ns + N_{sram} \times P_{sram} \quad (1)$$

Where $I_{src}$ is the current supplied by the voltage source, $N_{sram}$ is the number of configuration bits in the DUT, and $P_{sram}$ is the static power of the AOS SRAM cell at $V_{SRAM-CSB}$. Parameters defining the interfacial widths between bus connections and bus dimensionality are swept as CB and SB blocks steer connectivity through/off the routing bus interface (Fig. 6b). For reference, a single SB-Mux or CB-Mux refers to a modular building block of an SB or CB, in which a tile contains multiplicities of $W$ (× SB-Mux) and $I$ (× CB-Mux) respectively. We observe that, on average, switching to AOS pass gates and overdriving $V_{SRAM\text{-}SCB}$ to 1.2 V reduces the power consumption of SBs by ~13.7% and CBs by ~26% over CMOS at $V_{SRAM}$ of 0.8 V, closing the gap between ASIC and FPGA power efficiency.

## V. EVALUATION METHODOLOGY

### A. M3D Architectural Simulation of Xilinx-7 FPGA

Fig. 7 visualizes this work's evaluation flow of devices, circuits, and architectural measurement. To deal with the novel M3D placement structure and optimization of AOS PG widths and to develop architectural level parameters of our design, we develop a custom version of the tool COFFE [15] that can interface with the latest version of the academic Verilog-to-Routing (VTR) suite that is widely used for FPGA architectural benchmarking [16]. Circuit simulations are carried out in Synopsys HSPICE and Cadence Virtuoso using the compact device models described in prior sections and the ASAP7 PDK. Cost optimization is exhaustively performed for benchmarked circuits and tiles using the area-delay product. The Xilinx-7 architectural configuration file from VTR 8 is modified using M3D COFFE to reflect the interconnect, area, and delay changes in 7 nm technology and using the substitution of AOS transistors. A custom version of NeuroSim [33] is employed to estimate PPA parameters in a 32kB 7 nm FinFET dual-bank block RAM. VPR is used for CPD, total logic area utilization ($A_{TOT}$), maximum operation frequency ($F_{MAX}$), and $AT^2$.

### B. Benchmarking for Hyperdimensional Computing & LLMs

The growing need for efficient processing of high-dimensional data and long unstructured text warrants the study of hyperdimensional computing and large language model (LLM) acceleration on FPGAs to leverage their parallelism and energy efficiency (Table II). To benchmark performance on hyperdimensional computing, we select FFT [34], CNN [35], and GEMM [36] for their widespread use in digital signal processing and machine learning. Similarly, SPMV, characterized by its memory-bound operations, aligns well with FPGA architectures, permitting more energy-efficient and higher-performance execution compared to GPUs or other general-purpose processors; numerous efforts have been made to enhance its performance on FPGAs. AES [37] and Huffman Coding [38], essential for data encryption and compression, involve multiple sub-kernels and loop iterations. These benchmarks measure an FPGA's capability to handle compute-intensive, memory-bound tasks and complex control flows.

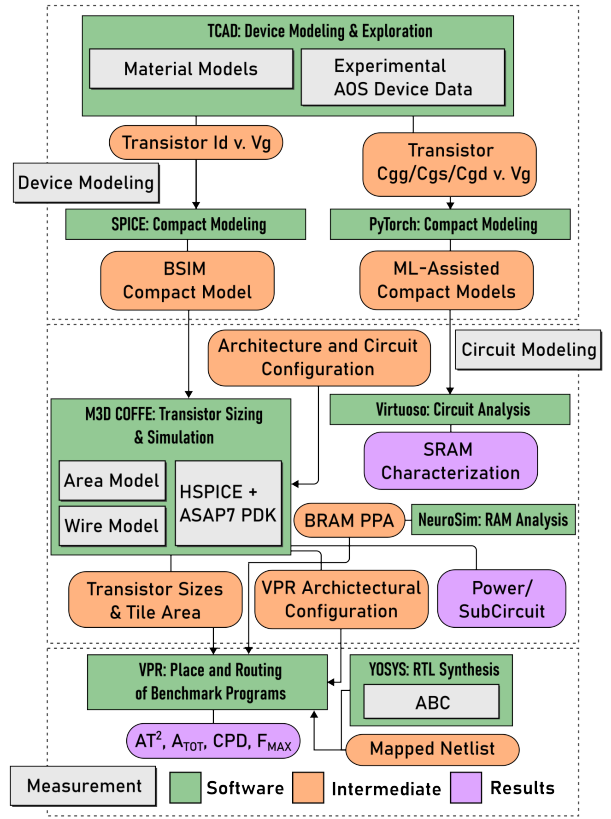

Fig. 7. Benchmarking methodology of 7 nm CMOS and M3D FPGAs

TABLE II
Hyperdimensional Computing Benchmark Setup

| Benchmark | Parameters | Input Type |
|---|---|---|
| FFT | Input data size n=64 | 32-bit FP |
| Convolution | Input size =128×228×228<br>Kernel size = 128×128×5×5<br>Output size = 128×224×224<br>5×5 convolution unit num = 16 | 8-bit int |
| GEMM | Matrix size=512×512<br>Systolic array size=8×8 | 8-bit int |
| SPMV | Single HiSparse [39] SPMV Cluster | 32-bit FP |
| AES Encoder & Decoder | Key size=256-bit | 128-bit int |
| Huffman Encoder | Input symbol size = 256<br>Max tree-construction depth = 64<br>Max rebalanced tree depth = 27 | 10-bit uint<br>32-bit uint |

## VI. BENCHMARKING RESULTS

### A. Hyperdimensional Computing Benchmark Comparison

Comparisons of the system level CPD, $A_{TOT}$, and $AT^2$ and the geometric mean for benchmarks in Table II are shown in Fig. 8, derived from VPR. We assess two systems in our study, both based on the Xilinx-7 architecture. The first is a baseline conventional CMOS design calibrated using 7 nm FinFET models from ASAP7 using COFFE with rails $V_{DD}$=0.7 V, $V_{SRAM}$=0.8 V. The second is our M3D FPGA design with AOS SRAM and PG substitution, which is calibrated in M3D COFFE utilizing our AOS BSIM-CMG model and ASAP7 with rails $V_{DD}$=0.7 V, $V_{SRAM}$=0.8 V and $V_{SRAM-SCB}$=1.2 V. Benchmarking findings indicate that our M3D FPGA design can decrease $A_{TOT}$ by >52%, CPD by 11.9% - 30%, and $AT^2$ by 63.6% - 77.1%. From prior works, we estimate that this reduction in performance and area would close the gap between ASICs and FPGAs down to 2.05-4.37× and 4.2× based on findings in [5].

Additionally, in Fig. 9a, we observe changes to the average distribution of routing utilization across relevant benchmarks at different localized and unlocalized segment lengths. In the M3D design, we find that the optimized layout post P&R has a higher favorability for longer routing segments due to the reduced cost of global routes. By translation, as is seen in Fig. 9b, which visualizes the congestion and critical path map from the AES benchmark, the congestion of nets becomes far more distributed on-chip. This can help delocalize hotspots and improve the placement proximity of hard macros in FPGAs.

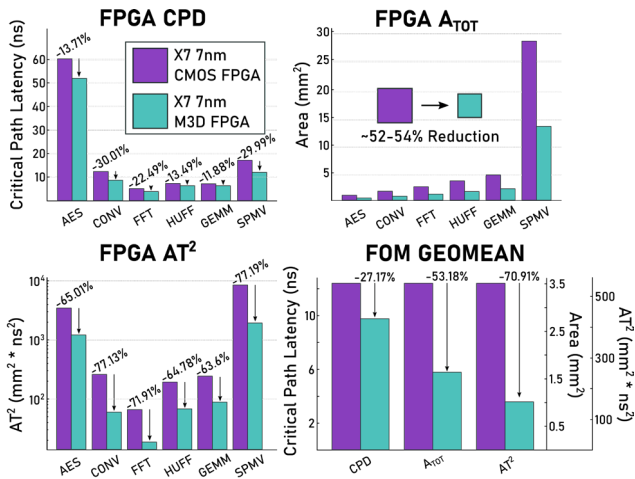

Fig. 8. Comparison of critical path delay, cumulative logic area, $AT^2$, and geometric mean across hyperdimensional computing benchmarks

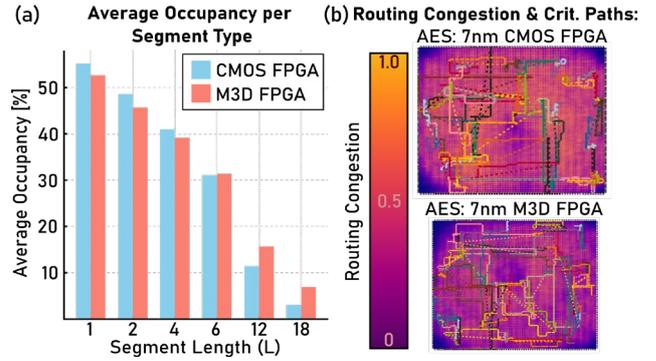

Fig. 9. (a) Comparison of routing occupancy, (b) congestion and critical paths across benchmarks in Xilinx-7 based CMOS and M3D FPGAs. M3D has a higher affinity for longer segments, reducing the signal congestion

### B. Towards GPT-Based LLM Acceleration

Given the rapid growth of LLM parameter sizes, we target a GPT-2 medium model accelerator implemented on the AMD Xilinx Versal VPK180 FPGA and estimate the utilization area based on the resource allocation relative to the total chip area of the VPK180. Delay is calculated based on the latency of critical paths after placement and routing using Vivado 2023.2. Aiming to transfer the same design to an M3D FPGA, we estimate the area and delay, assuming identical resource utilization. We observe a 63.7% improvement in $AT^2$ when translating the design from the VPK180 to an M3D equivalent, achieving significant reductions in both area and delay (Table III).

TABLE III
Comparison of LLM (GPT-2) Implementation on M3D FPGA

| System | Versal VPK 180 | 7nm M3D FPGA |
|---|---|---|
| Area | 736.0 mm² | 423.8 mm² (-42.4%) |
| Delay | 4.17 ns | 3.31 ns (-20.6%) |
| $AT^2$ | 12798.9 mm²×ns² | 4643.2 mm²×ns² (-63.7%) |

## VII. CONCLUSIONS

This work presents a design-space analysis of a novel M3D FPGA architecture with AOS device-based configuration SRAMs and routing pass gate structures used in switch and connection blocks. A CAD flow for exploring the proposed M3D design that can optimize the placement and sizing of AOS devices is developed to evaluate performance on a suite of compute-intensive hyperdimensional computing tasks ranging from 170×170 to 372×372 logical blocks per net. The proposed M3D FPGA demonstrates improvements of up to 30% lower delay, 54% lower system area, and 77% lower $AT^2$. By transitioning to AOS devices with reduced static power, monolithically stacking devices to create multiple active tiers, and decreasing the overhead of interconnects, the proposed design cuts the PPA disparity between ASICs and FPGAs, improving the utility of FPGAs in data-intensive applications such as hyperdimensional computing and LLM acceleration.

## VIII. Acknowledgments

This work is supported by PRISM, an SRC/DARPA JUMP 2.0 center. The authors thank S. Jung from Georgia Tech and S. More from Univ. of Toronto for guidance on device modeling and open-source FPGA EDA tools.